\documentclass[useAMS,usenatbib]{mn2e}

\usepackage{graphicx}
\usepackage{txfonts}

\title[Global abundance determinations]
      {Abundance determination from global emission-line SDSS spectra:
       exploring objects with high N/O ratios}

\author[L.S.Pilyugin et al.]
       {L.S.~Pilyugin$^{1,2}$,
        J.M.~V\'{\i}lchez$^{2}$, 
        L.~Mattsson$^{3}$, 
        T.X. Thuan$^{4}$ \\
     $^{1}$ Main Astronomical Observatory
            of National Academy of Sciences of Ukraine,
            27 Zabolotnogo str., 03680 Kiev, Ukraine \\
     $^{2}$ Instituto de Astrof\'{\i}sica de Andaluc\'{\i}a,
            CSIC, Apdo, 3004, 18080 Granada, Spain \\
     $^{3}$ Dark Cosmology Centre, Niels Bohr Institute,
            University of Copenhagen, Juliane Maries Vej 30,
            DK-2100, Copenhagen \O, Denmark\\
     $^{4}$ Astronomy Department, University of Virginia, 
            P.O. Box 400325, Charlottesville, VA 22904-4325, USA \\
              }

\date{Accepted 2011 December 20. Received 2011 December 19; in original form 2011 November 17}

\begin{document}

\maketitle

\begin{abstract}
We have compared the oxygen and nitrogen abundances derived from global 
emission-line SDSS spectra of galaxies using (1) the $T_e$ method and (2) 
two recent strong line calibrations: the ON and NS calibrations. 
Using the $T_e$ method, anomously high N/O abundances ratios have been found 
in some SDSS galaxies. To investigate this, we have  Monte Carlo simulated the global 
spectra of composite nebulae by a mix of spectra of individual components, based on spectra 
of well-studied H\,{\sc ii} regions in nearby galaxies. We found that the $T_e$ method 
results in an underestimated oxygen abundance (and hence in an overestimated 
nitrogen-to-oxygen ratio)  if H\,{\sc ii} regions with different physical properties contribute 
to the global spectrum of composite nebulae. This effect is somewhat similar 
to the small-scale temperature fluctuations in H\,{\sc ii} regions discussed by Peimbert. 
Our work thus suggests that the high $T_e$-based N/O abundances ratios 
found in SDSS galaxies may not be real. However, such an effect is not expected to be 
present in dwarf  galaxies since they have generally an uniform chemical composition.
The ON and NS calibrations give O and N abundances in composite nebulae 
which agree with the mean luminosity-weighted abundances of their components 
to within  $\sim$0.2 dex. 
\end{abstract}


%________________________________________________________________

\section{Introduction}

Metallicities play a key role in many studies of galaxies.
Gas-phase oxygen and nitrogen abundances are broadly used to estimate 
these metallicities. It is believed \citep[e.g.][]{stasinska2006} that 
emission lines in spectra of H\,{\sc ii} regions are the most powerful 
indicators of the chemical composition of galaxies, both in the low- and 
intermediate-redshift universe. 
The spectra of a large number of individual H\,{\sc ii} regions in nearby 
spiral and irregular galaxies have now been obtained 
\citep[][among many others]{mccalletal1985,zaritskyetal1994,vanzeeetal1998,
vanzeehaynes2006,bresolinetal1999,bresolinetal2005,bresolinetal2009}. 
These spectroscopic 
measurements provide basis for investigations of metallicity 
properties (such as radial abundance gradients, central metallicities, etc) of 
galaxies \citep[][among others]{zaritskyetal1994,vanzeeetal1998,
bresolinetal1999,pilyuginetal2004b}.

\citet{kennicutt1992} pioneered another method of spectral investigation of 
galaxies, considering the global, i.e. spatially unresolved, spectrophotometry of a
sample of 90 galaxies spanning the entire Hubble sequence. 
In recent years, the number of available spectra of emission-line 
nebulae has increased dramatically due to the completion of several 
large spectral surveys such as 
the Sloan Digital Sky Survey \citep[SDSS,][]{yorketal2000}. 
Measurements of emission lines in SDSS spectra have been used for 
abundance determinations  in a number of studies
\citep[][among others]{kniazevetal2004,izotovetal2006,tremontietal2004,
asarietal2007,thuanetal2010}. 
The auroral lines are measurable in a relatively large number of SDSS galaxies 
\citep{kniazevetal2004,izotovetal2006} that provide the possibility to obtain 
$T_e$-based abundances for SDSS galaxies.

The SDSS spectra are obtained through 3-arcsec diameter fibers. At a redshift of $z=0.025$, 
the projected aperture diameter is $\sim$ 1.5 kpc, while it is  $\sim$ 10 kpc at a redshift of $z=0.17$. 
This suggests that SDSS spectra of distant galaxies are closer to global spectra of galaxies, i.e. they 
are the integrated spectra of  multiple rather than just one H\,{\sc ii} region. Therefore the meaning of   
$T_e$-based abundances in SDSS galaxies is unclear. If a single giant H\,{\sc ii} region, excited by a 
single star cluster, is responsible for the global SDSS spectrum then one can expect that the $T_e$-based 
abundance to be a good estimate of the true one. Such a situation occurs only in nearby SDSS galaxies 
and in some distant SDSS galaxies where a single supergiant H\,{\sc ii} region, resulting from a strong 
starburst, makes a dominant contribution to the global spectrum. One can expect, however, that in a 
majority of distant SDSS galaxies, multiple individual H\,{\sc ii} regions contribute to the global spectrum. 

\citet{ercolanoetal2007,ercolanoetal2010} have investigated the effect of multiple 
ionization sources in  H\,{\sc ii} regions on the total elemental abundances derived 
from the analysis of collisionally excited emission lines. They 
considered the case of an ionizing set composed of two stellar 
populations with masses of 37 $M_{\sun}$ and 56 $M_{\sun}$. 
They have shown that the temperature structures of models 
with centrally concentrated ionizing sources can be quite different  
from those of models where the ionizing sources are randomly distributed within 
the volume, with generally non-overlapping Str\"omgren spheres.  
Since the SDSS composite nebula can be ionised by sources of different temperatures   
then the meaning of the measured electron temperature in distant SDSS galaxies 
and consequently, that of the $T_e$-based abundance is not evident. 

This problem is in some sense similar to the one concerning the validity of the $T_e$-based 
abundances in H\,{\sc ii} regions with small-scale temperature fluctuations \citep{peimbert1967} 
or/and with large variations of the electron temperature across the nebula 
\citep{stasinska1978,stasinska2005}. According to Peimbert, since the line fluxes in the  spectrum  
of such a nebula are weighted in favor of  the hot regions, the electron temperature estimated 
from the auroral-to-nebular lines ratio will be overestimated and the abundance underestimated. 
While the existence of such temperature fluctuations in individual H\,{\sc ii} regions 
is still debated, temperature variations of individual H\,{\sc ii} regions  
within a composite nebula are to be  perfectly natural.
Thus, such an effect is probably at work in the distant SDSS galaxies. 

When the electron temperature of an extragalactic H\,{\sc ii} region cannot be measured, then its location
in some emission-line diagrams is used for estimating its oxygen 
abundance. This approach to abundance determination 
in H\,{\sc ii} regions, proposed by \citet{pageletal1979} and 
\citet{alloinetal1979}, is usually referred to as the ``strong line method''. 
Numerous relations have been proposed to convert metallicity-sensitive emission-line 
ratios into metallicity or temperature estimates
\citep[e.g.][]{dopitaevans1986,zaritskyetal1994,vilchezesteban1996,pilyugin2000,pilyugin2001,
denicoloetal2002,pettinipagel2004,tremontietal2004,pilyuginthuan2005,
liangetal2006,stasinska2006,bresolin2007,perezmontero2009,thuanetal2010}. 
Although SDSS spectra of distant galaxies are closer to global galaxy 
spectra than to spectra of individual H\,{\sc ii} regions, 
abundances in distant SDSS galaxies have been estimated using the strong line 
methods developed for abundance determination in individual H\,{\sc ii} regions 
\citep[][among others]{tremontietal2004,erbetal2006,asarietal2007,thuanetal2010}. 
This, despite the warning of \citet{stasinska2010} that the strong line 
methods should be used only for nebulae having the same structural 
properties as those of the calibration samples. 

However, there is a reason to expect the calibrations may provide more robust abundances in 
composite nebulae than the $T_e$  method. For definiteness, we will discuss the case of oxygen 
abundances. The auroral and nebular lines originate in transitions from levels 
that differ significantly in energy, with the levels that give rise to the  
auroral line being at higher energies than those for nebular lines. Therefore, the  emissivity
of the auroral line depends much more strongly on the electron temperature in the 
nebula than that of the nebular line.  Hence, the auroral and nebular line fluxes 
in the integrated spectra of composite nebulae are weighted in different ways: 
the hot regions will give relatively more weight to the auroral line than 
to the nebular lines. This means the nebular and auroral lines in the spectra of composite 
nebulae correspond to different temperatures, i.e. they are not  self-consistent. The electron 
temperature estimated from the auroral-to-nebular lines ratio will then be overestimated 
and the abundance will be underestimated \citep{peimbert1967,stasinska1978,stasinska2005}.   
In the case of the empirical calibrations considered here (the ON and NS calibrations), 
combinations of strong lines for different ions serve as abundance and electron 
temperature indicators. 
The differences between the energies of parent levels for these nebular lines 
are smaller than those between the energies of parent levels for 
the auroral and nebular lines. Then one can expect that, in the strong line method, the different 
regions will give roughly similar relative contributions to the nebular 
line fluxes, i.e. they will all correspond 
to more or less similar electron temperatures. 
This, in turn, will result in more robust global abundances estimates.

The goal of the present study is to examine to what extent abundances 
for distant SDSS galaxies derived in different ways (by the classic 
$T_e$ method and through recent calibrations) agree or disagree.
The abundances of a sample of SDSS galaxies are examined in Section 2. 
In Section 3, we compare the data with artificial global spectra of composite nebulae
generated by Monte Carlo techniques 
as mixes of spectra of individual components, using spectra of real  H\,{\sc ii} 
regions in nearby galaxies with measured electron temperatures.  
The meaning of abundances 
derived in composite nebulae in different ways and their  
locations in the O/H -- N/O 
diagram are examined. A discussion of the results is given in Section 4.
Section 5 presents the conclusions. 

%====================================       Fig  No 1   logR3  -  logR2
\begin{figure}
\resizebox{0.95\hsize}{!}{\includegraphics[angle=000]{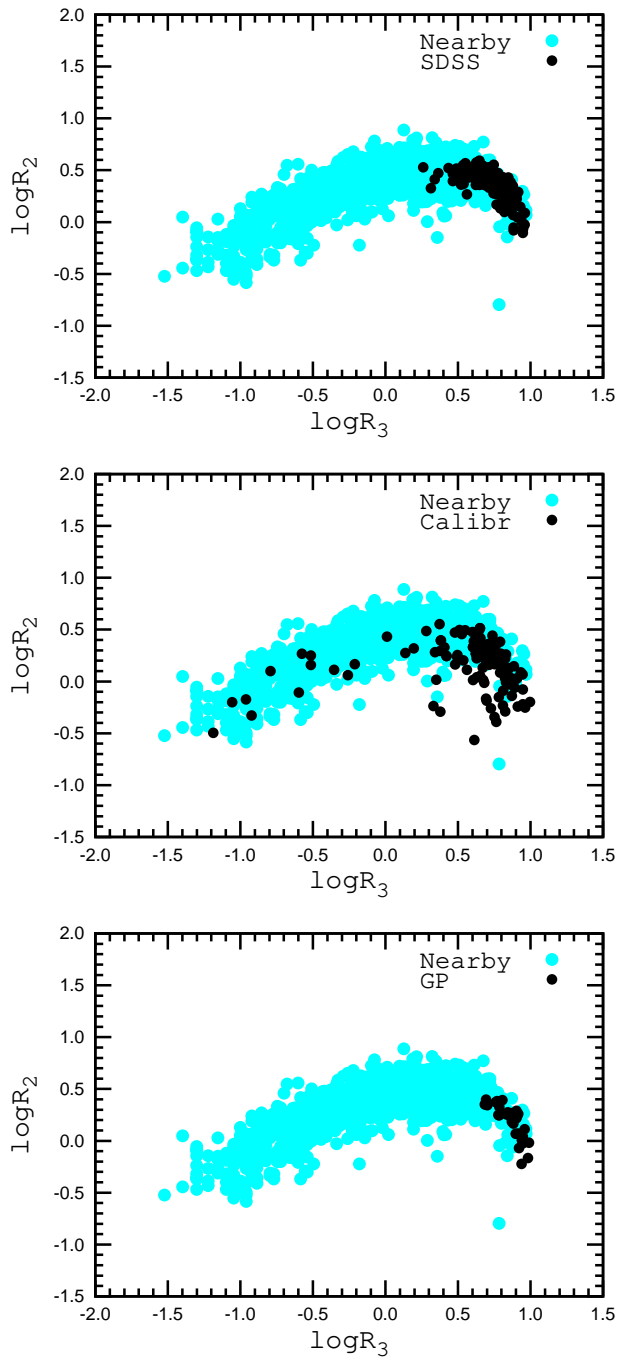}}
\caption{ The log$R_3$ -- log$R_2$ diagram. 
The filled dark (black in the color version) circles 
show the objects from our SDSS sample (upper panel), the sample of calibration 
H\,{\sc ii} regions (middle panel), and the Green Pea galaxies (lower panel).
The filled grey (light-blue in the color version) circles in each panel are 
objects from the sample of H\,{\sc ii} regions in nearby galaxies.
(A color version of this figure is available in the online version.)
}
\label{figure:lr3-lr2}
\end{figure}

Throughout the paper, we will be using the following standard notations for the line 
intensities: \\ 
$R$  = $I_{[OIII] \lambda 4363} /I_{{\rm H}\beta }$,  \\
$R_2$  = $I_{[OII] \lambda 3727+ \lambda 3729} /I_{{\rm H}\beta }$,  \\
$N_2$  = $I_{[NII] \lambda 6548+ \lambda 6584} /I_{{\rm H}\beta }$,  \\
$S_2$  = $I_{[SII] \lambda 6717+ \lambda 6731} /I_{{\rm H}\beta }$,  \\
$R_3$  = $I_{{\rm [OIII]} \lambda 4959+ \lambda 5007} /I_{{\rm H}\beta }$.  \\
The electron temperatures will be given in units of 10$^4$K.

\section{Oxygen and nitrogen abundances}
%================

\subsection{Galaxy samples}
%================

\subsubsection{SDSS galaxies}
In our previous work \citep{pilyuginetal2010a},
we have constructed a sample of galaxies by selecting from Data Release 6 
of the SDSS the galaxy spectra which satisfy the following criteria: 
(1) both [O\,{\sc iii}]$\lambda$4363 
and [O\,{\sc ii}]$\lambda$7320+$\lambda$7330 auroral lines are detected;  
(2) the spectra have smooth line profiles. Particular attention was paid to 
the two auroral lines mentioned above   
since the accuracy of the electron 
temperature determination depends mainly  
on those emission lines; 
(3) the emission lines do not have a broad component. 
The line intensities in these spectra have been measured by  
fitting every line with a Gaussian profile and de-reddened in the way described in 
\citet{pilyuginthuan2007,pilyuginetal2010a}. 

The wavelength range of the SDSS spectra is 3800 -- 9300\AA$.$ Hence,
for nearby galaxies with redshift $z$ $\la$ 0.02, the 
[O\,{\sc ii}]$\lambda$3727+$\lambda$3729 emission line is outside that range. 
The absence of this line prevents determination of the 
oxygen abundance through the standard version of the $T_e$ method. We thus 
exclude the most nearby galaxies.
All galaxies in our sample have redshifts larger than 
$\sim$ 0.023, i.e. they are more distant than $\sim$ 100 Mpc.
We thus obtain a total sample of 281 SDSS galaxies. 
The upper panel of Fig.~\ref{figure:lr3-lr2} shows the positions of the SDSS galaxies of
our sample in the standard $R_{3}$ --  $R_{2}$ diagram by 
filled dark (black in the color version) circles. 

\subsubsection{H\,{\sc ii} regions in nearby galaxies}
\citet{pilyuginetal2004b} have made a compilation of a 
large amount of strong emission line measurements in spectra of individual 
H\,{\sc ii} regions in nearby spiral and irregular galaxies. 
Their sample consists of 1121 data points and will be used to outline the area occupied by  
 H\,{\sc ii} regions of nearby galaxies in the $R_{3}$ --  $R_{2}$ diagram. 
These are shown by filled grey (light-blue in the color version) circles in the 
$R_{3}$ --  $R_{2}$ diagram. 
Fig.~\ref{figure:lr3-lr2} shows that the SDSS galaxies are located in a relatively 
small part of the area occupied by H\,{\sc ii} regions in nearby galaxies. 

\subsubsection{Calibration H\,{\sc ii} regions}
The [O\,{\sc iii}]$\lambda$4363 or/and [N\,{\sc ii}]$\lambda$5755  auroral lines 
are detected in many  H\,{\sc ii} regions in nearby galaxies. A compilation of such  
H\,{\sc ii} regions was made by  \citet{pilyugin2010ons} and used 
as calibration data points to derive the 
relations which give the oxygen and nitrogen abundances and electron temperature  
in terms of the strong emission line fluxes. 
After the exclusion of a few ``peculiar'' objects  \citep{pilyugin2010ons}, we end up 
with 112 data points.
The spectra of this sample of H\,{\sc ii} regions in nearby spiral and irregular galaxies 
with detected auroral lines will be used as templates to simulate artificial spectra of composite 
nebulae. The middle panel of Fig.~\ref{figure:lr3-lr2} shows the positions of the calibration 
 H\,{\sc ii} regions in the  $R_{3}$ --  $R_{2}$ diagram as filled dark (black) circles. The filled 
 grey (light-blue) circles are the same as in the upper panel. 

\subsubsection{Green Pea galaxies}
The Green Pea galaxies have been extracted from the SDSS sample 
by \citet{cardamone2009}. 
These galaxies have a compact appearance and are characterised by distinct green 
colours in $gri$ images. 
\citet{cardamone2009} found that the Green Peas are low-mass galaxies 
($M \sim 10^{8.5}$ -- $10^{10} M_{\sun}$) with high star formation rates  
($\sim  10 M_{\sun}$ yr$^{-1}$) with metallicities 12 +log(O/H) $\sim$  8.7. 
\citet{izotovetal2011} have argued that they form a subset of a larger 
population of luminous compact star-forming galaxies. 
 \citet{amorin2010} have derived $T_e$--based oxygen abundances 
 in these galaxies and have 
found that the Green Pea galaxies show metallicities  7.7 $\la$ 12 + log(O/H) $\la$ 8.4, 
with a mean value of 8.05 $\pm$ 0.14. They have also found that some  
Green Pea galaxies display enhanced N/O ratios. 
\citet{izotovetal2011} have found that the oxygen abundances 12 + log(O/H) in 
 luminous compact star-forming galaxies are in the range 7.6 -- 8.4. They found no 
appreciable difference in element abundances between  
luminous compact star-forming galaxies 
and local blue compact dwarf galaxies, implying a similar chemical enrichment history.

From the sample of Green Pea galaxies of \citet{cardamone2009} we have selected
those with detectable [O\,{\sc iii}]$\lambda$4363 auroral lines.  
The line fluxes in the SDSS spectra were measured 
with {\sc iraf}  \footnote{{\sc iraf} is distributed by National Optical Astronomical Observatories,
which are operated by the Association of Universities for Research in Astronomy, Inc., 
under cooperative agreement with the National Science Foundation.}.
The measured line fluxes were dereddened in the way described in 
\citet{pilyuginthuan2007,pilyuginetal2010a}.
The lower panel of Fig.~\ref{figure:lr3-lr2} shows the positions of the Green Pea galaxies 
in the $\log R_3$ -- $\log R_2$ diagram (filled dark (black) circles).  
 The filled  grey (light-blue) circles are the same as in the upper panel.
Note that the Green Pea galaxies overlap with the area 
occupied by our SDSS sample in the $R_{3}$ --  $R_{2}$ diagram.

\subsection{Abundance determination}
%-----------------------------------

Line fluxes are converted to electron temperatures 
and ion abundances within using the standard 
H\,{\sc ii} region model with two distinct temperature zones within 
the nebula:  the electron temperatures $t_2$ within 
the O$^+$ zone and $t_3$ within the O$^{++}$ zone. 
Electron temperatures and ion 
abundances from line fluxes were derived as according to \citet{pilyugin2010ons}.
The ratio of the nebular to auroral oxygen line intensities 
[O\,{\sc iii}]$\lambda 4959+\lambda 5007$/[O\,{\sc iii}]$\lambda 4363$ 
is used for the $t_{3}$ determination and 
the ratio of the nebular to auroral nitrogen line intensities 
[N\,{\sc ii}]$\lambda 6548+\lambda 6584$/[N\,{\sc ii}]$\lambda 5755$ 
is used for the $t_{2}$ determination.

When only one of the electron temperatures is known, 
it is common practice to estimate the other temperature by 
using a relation between $t_2$ and $t_3$. 
We have adopted the following $t_2$--$t_3$ relation 
\begin{equation}
t_2 = 0.672\,t_3 + 0.314 .
\label{equation:t2t3}   
\end{equation}
This relation derived by \citet{pilyuginetal2009} is very similar to the widely  
used one proposed by \citet{campbell1986} and confirmed by \citet{garnett1992}.
In general, using the classic $T_e$ method, one can derive the oxygen 
abundance in each H\,{\sc ii} region in several ways: 
(O/H)$_{t_3}$ abundances can be determined with the 
measured $t_3$, $t_2$ being then estimated from the $t_2$--$t_3$ relation; or  
(O/H)$_{t_2}$ abundances can be found with a measured $t_2$,  
$t_3$ being then estimated from the $t_2$--$t_3$ relation. 
If the line-flux measurements 
are accurate enough, the oxygen abundances derived in these two  
ways should agree. 
The (O/H)$_{t_3}$ abundance will be considered here and will be 
referred to as (O/H)$_{T_e}$ hereafter. 

We also estimate the oxygen and nitrogen abundances using two recent 
empirical calibrations. 
The ON-calibration relations give the oxygen and nitrogen abundances and electron temperature  
in terms of the fluxes of the strong emission lines O$^{++}$, O$^{+}$, and 
N$^+$. It has been derived using spectra of H\,{\sc ii} regions with well-measured electron temperatures 
as calibration datapoints \citep{pilyugin2010ons}.  
The oxygen and nitrogen abundances estimated using the ON calibration will be referred to below as 
(O/H)$_{ON}$ and (N/H)$_{ON}$.

In addition to the direct method and the ON-calibration, the NS-calibration is used as well to estimate oxygen 
and nitrogen abundances. These relations give abundances and electron temperatures 
in terms of the fluxes in the strong emission lines of O$^{++}$,
N$^+$, and S$^+$, and have also been derived using spectra of
H\,{\sc ii} regions with well-measured electron temperatures as calibration datapoints 
\citep{pilyuginmattsson2011}. 
The oxygen and nitrogen abundances estimated using the NS calibration will be referred to below as 
(O/H)$_{NS}$ and (N/H)$_{NS}$.

%====================================       Fig  No 2   N/O --O/H relation 
\begin{figure}
\resizebox{1.00\hsize}{!}{\includegraphics[angle=000]{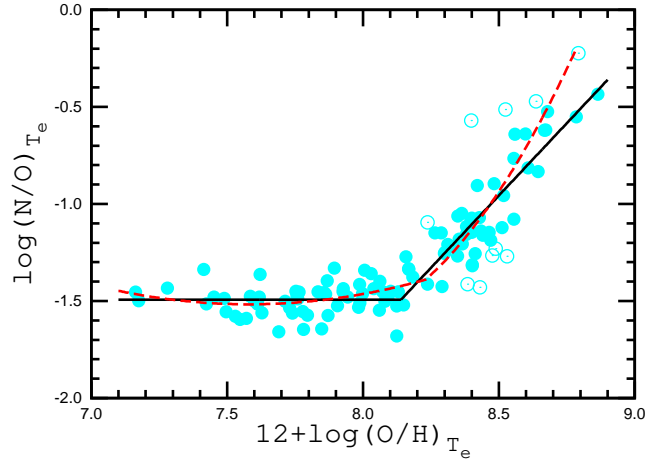}}
\caption{
The O/H -- N/O diagram for the calibration H\,{\sc ii} regions. 
The solid (black) line shows a linear fit to the N/O -- O/H relation 
(Eq.~\ref{equation:nooh-relat}). 
The filled circles are data points used in deriving this relation, 
the open circles are rejected data points. 
The dashed (red) line shows an alternative quadratic fit to the 
N/O -- O/H relation. 
}
\label{figure:oh-no-relat} 
\end{figure}

\subsection{The O/H -- N/O diagram}
%----------------------------------

Fig.~\ref{figure:oh-no-relat} shows a standard O/H -- N/O diagram for 
 the calibration H\,{\sc ii} regions. 
Using these data, we have obtained the following linear 
relations between log N/O and log O/H   
\begin{eqnarray}
       \begin{array}{lll}
\log {\rm (N/O)}  & =  &  -1.493                                            \\
                  &    & {\rm for} \quad 12+\log {\rm (O/H)} < 8.14  ,      \\
                  &    &                                                    \\
                  & =  &  1.489 \times (12+\log {\rm (O/H))} - 13.613       \\
                  &    & {\rm for} \quad 12+\log {\rm (O/H)} > 8.14  .      \\
     \end{array}
\label{equation:nooh-relat}
\end{eqnarray}
Several H\,{\sc ii} regions show large ($>$0.25 dex) abundance deviations. These have been excluded 
in deriving the final relation. 
In Fig.~\ref{figure:oh-no-relat}, the data points used in the derivation of the final relation 
(102 out of 112) are shown by filled circles, while   
the rejected data points are shown by open circles.  
The solid (black) line shows the derived relation.
An alternate quadratic relation between log N/O and log O/H has 
also been derived: 
\begin{eqnarray}
       \begin{array}{lll}
\log {\rm (N/O)}  & =  &  -1.39 + 0.39\, (Z - Z_0) + 0.30 \, (Z - Z_0)^2    \\
                  &    & {\rm for} \quad Z  < 8.23         ,              \\
                  &    &                                                    \\
                  & =  &  -1.39 + 1.24\, (Z - Z_0) + 1.63 \, (Z - Z_0)^2    \\
                  &    & {\rm for} \quad Z  > 8.23                 ,      \\
     \end{array}
\label{equation:nooh-relat2}
\end{eqnarray}
where $Z$ = 12 + log(O/H) and  $Z_0$ = 8.23 (see the dashed (red) line in Fig.~\ref{figure:oh-no-relat}).
The two relations are similar for metallicities  12 + log(O/H) $\la$ 8.5. We will here only consider
objects in this metallicity range. 
Using the N/O -- O/H relationship, one can estimate the N/O ratio which corresponds 
to a given oxygen abundance. The nitrogen-to-oxygen abundance ratio obtained in this way will be 
referred to as (N/O)$_{REL}$ below.
For definiteness, we will use the N/O -- O/H 
relationship given by Eq.~\ref{equation:nooh-relat}.

%====================================       Fig  No 3   observed O/H - N/O  diagram  
\begin{figure}
\resizebox{0.90\hsize}{!}{\includegraphics[angle=000]{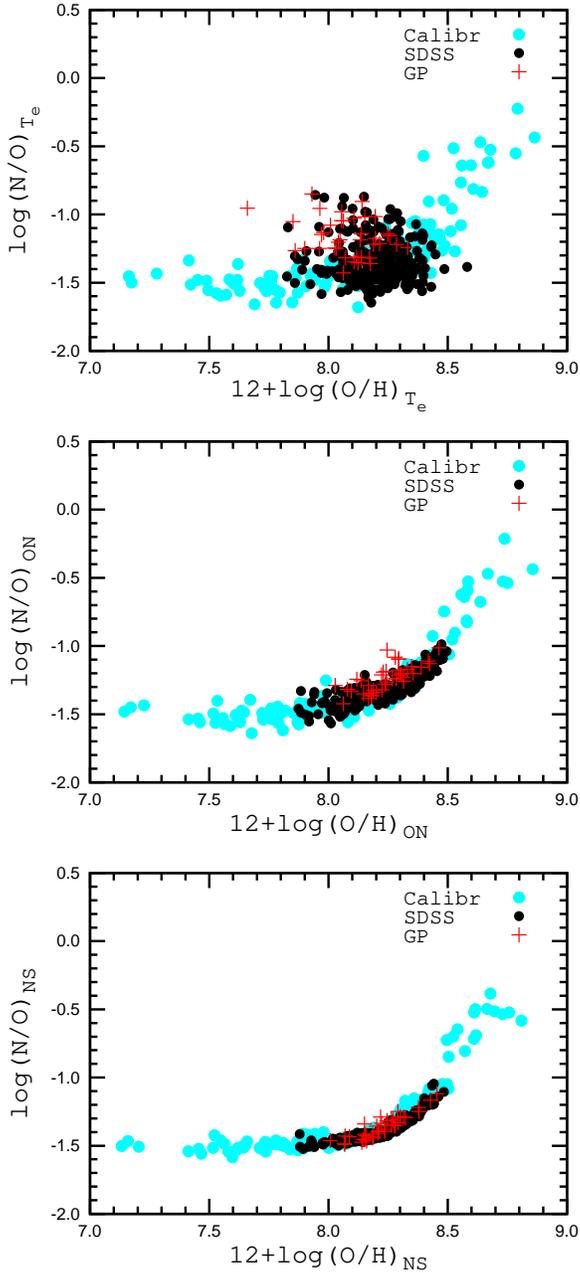}}
\caption{
The O/H -- N/O diagrams for SDSS objects (filled dark 
(black) circles), for the Green Pea galaxies (dark (red) plus signs), 
and for calibration H\,{\sc ii} regions (filled grey (light-blue) circles) 
with abundances obtained with the $T_e$ method (upper panel), 
using the ON calibration (middle panel), and the NS calibration (lower panel). 
 (A color version of this figure is available in the online version.)
}
\label{figure:ohno-s} 
\end{figure}

We have derived (O/H)$_{T_e}$,  (O/H)$_{ON}$,  (O/H)$_{NS}$ oxygen abundances 
and (N/H)$_{T_e}$,  (N/H)$_{ON}$,  and (N/H)$_{NS}$ nitrogen abundances 
for all calibration H\,{\sc ii} regions, SDSS and Green Pea galaxies.  
Fig.~\ref{figure:ohno-s} shows the comparison between O/H -- N/O diagrams 
for abundances computed in different ways.

The upper panel of Fig.~\ref{figure:ohno-s} shows the O/H -- N/O diagram 
for SDSS objects 
(filled dark (black) circles), Green Pea galaxies (dark (red) plus signs), and 
 calibration H\,{\sc ii} regions (filled grey (light-blue) circles)
for abundances obtained with the $T_e$ method. 
The upper panel shows that some SDSS objects deviate significantly
from the general trend defined by the calibration H\,{\sc ii} regions:  
they are shifted towards higher N/O 
abundances ratio or/and towards lower O/H abundances. The same shift 
is observed for some Green Pea galaxies, 
in agreement with the findings of \citet{amorin2010}.

The middle panel of Fig.~\ref{figure:ohno-s} shows the  same O/H -- N/O diagram, but
for the case where abundances are obtained with the ON calibration, and shows that the SDSS 
and Green Pea galaxies follow the trend defined by the calibration H\,{\sc ii} regions, 
without large deviations. 
The lower panel of Fig.~\ref{figure:ohno-s} shows the NS calibration-based abundances.  
Again,  the SDSS and Green Pea galaxies follow the trend defined by the  
calibration H\,{\sc ii} regions and do not show large deviations.

%====================================       Fig  No 4   dNO deviations 
\begin{figure}
\resizebox{0.90\hsize}{!}{\includegraphics[angle=000]{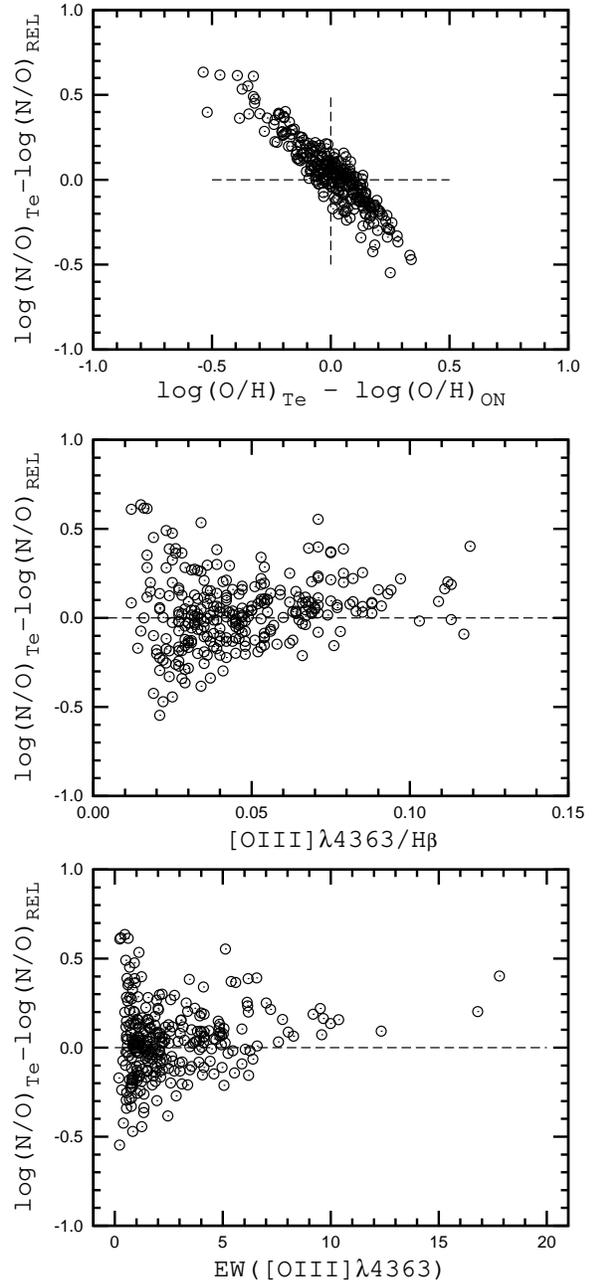}}
\caption{
The deviation of the N/O ratios in SDSS objects 
relative the N/O -- O/H relation  
as a function of (1) the difference between  $T_e$-based and ON-calibration-based 
oxygen abundances (upper panel), 
(2) of the flux in the auroral line  
[O\,{\sc iii}]$\lambda$4363 normalised to the H$\beta$ line flux (middle 
panel), and (3) of the equivalent width EW([O\,{\sc iii}]$\lambda$4363) of the 
auroral line (lower panel).
}
\label{figure:do-dno-s}
\end{figure}

The upper panel of Fig.~\ref{figure:do-dno-s} shows the deviation of the 
(N/O)$_{T_e}$ ratio in the SDSS objects  from the N/O -- O/H relation 
as a function of the difference between  $T_e$-based and ON-calibration-based 
oxygen abundances. There is a clear anti-correlation between 
(log(N/O)$_{T_e}$ -- log(N/O)$_{REL}$) and (log(O/H)$_{T_e}$ --log(O/H)$_{ON}$) values. 
This anti-correlation suggests the deviation of SDSS objects 
from the general N/O -- O/H trend in the upper panel of 
 Fig.~\ref{figure:ohno-s} is mainly caused by shifts 
towards lower O/H abundances, i.e. the (O/H)$_{T_e}$ abundances in these 
objects appear to be underestimated.

%====================================       Fig  No 5   OH - NO mean 
\begin{figure}
\resizebox{1.00\hsize}{!}{\includegraphics[angle=000]{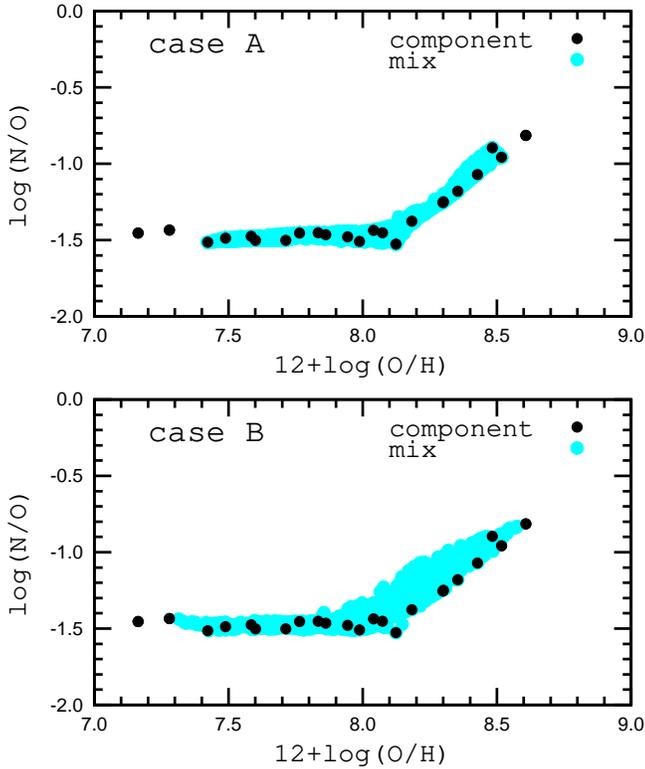}}
\caption{The O/H -- N/O diagram for artificial composite nebulae and components. 
{\it Upper panel.}
Filled dark (black) circles show the $T_e$-based abundances of components,
while the grey (light-blue) filled circles are abundances of artificial composite 
nebulae, calculated as the 
mean luminosity-weighted abundance of all components, for case $A$. 
{\it Lower panel.}
The same as in the upper panel, but for case $B$.
(A color version of this figure is available in the online version.)
}
\label{figure:mix-mean}
\end{figure}

One may expect the high (N/O)$_{T_e}$ ratios  
in SDSS objects to be caused by uncertainties in the flux of the weak 
auroral [O\,{\sc iii}]$\lambda$4363 line in these objects. To check this hypothesis, we 
show in the middle panel of Fig.~\ref{figure:do-dno-s} the deviation of (N/O)$_{T_e}$ ratios 
(log(N/O)$_{T_e}$ -- log(N/O)$_{REL}$) versus the flux of the  
[O\,{\sc iii}]$\lambda$4363 auroral line, normalised to the  H$\beta$ flux. 
The lower panel of Fig.~\ref{figure:do-dno-s} shows the deviation of the same 
(N/O)$_{T_e}$ ratios 
as a function of the equivalent width EW([O\,{\sc iii}]$\lambda$4363) of the  
[O\,{\sc iii}]$\lambda$4363 auroral line. 
Examination of these two panels suggests 
that the uncertainties in the O\,{\sc iii}]$\lambda$4363 line flux 
can indeed be responsible for a significant fraction of high (N/O)$_{T_e}$ 
ratios in 
the SDSS galaxies. However, high (N/O)$_{T_e}$ ratios
are found not only in SDSS objects with  a weak  [O\,{\sc iii}]$\lambda$4363 line, but also 
in some objects where that line is relatively strong, 
with EW([O\,{\sc iii}]$\lambda$4363) $>$ 5 and/or [O\,{\sc iii}]$\lambda$4363/H$\beta$ $>$ 0.05.  
This implies that the auroral line flux uncertainties cannot be the only 
reason for the high (N/O)$_{T_e}$ ratios found in some SDSS objects. 

We propose here another explanation for the derived high (N/O)$_{T_e}$ ratios. 
SDSS galaxies are composite nebulae which contain a number of  H\,{\sc ii} regions with 
different physical properties, all contributing to the global spectrum.  
In that case, the meaning of a single measured electron temperature for the 
whole galaxy is not well defined and consequently,  
$T_e$-based abundances are ambiguous. We expect the electron 
temperature estimated from the auroral-to-nebular lines ratio to be overestimated 
and the derived oxygen abundance to be underestimated. In other words, such nebulae 
will exhibit an effect similar to the 
temperature fluctuations discussed by \citet{peimbert1967}.  However, these 
temperature variations will not occur within a single H\,{\sc ii} region, but on the 
large scale, in ongoing from one H\,{\sc ii} region to other. 
These temperature variations in an object will cause it to deviate from the general 
N/O -- O/H trend.  To explore this scenario quantitatively, we use a 
Monte Carlo simulation to construct spectra of composite nebulae  
(described in the next section), and examine how derived abundances in composite nebulae 
change, depending on the abundance determination method.

\section{Global abundances in artificial composite nebulae}
%----------------------------------

\subsection{Monte Carlo simulation of global spectra}
%----------------------------------

To clarify the reason for the discrepancy between the $T_e$-based and 
calibration-based abundances in our sample of distant SDSS galaxies, 
we have simulated their spectra. 
We have remarked above that several different H\,{\sc ii} regions can 
contribute to the global spectra of distant SDSS galaxies.  
We have modelled their global spectra by using the approach 
of \citet{pilyuginetal2004a,pilyuginetal2010a}. 
The artificial spectra of SDSS galaxies have been computed as a mix of 
spectra of individual components, which we take to be spectra of real  H\,{\sc ii} 
regions in nearby galaxies with measured electron temperatures, taken from 
the calibration H\,{\sc ii} region sample.

The auroral line [O\,{\sc iii}]$\lambda$4363 is usually detected in spectra of low-metallicity 
calibration H\,{\sc ii} regions and $t_3$ can then be determined,  
while the  auroral line  [N\,{\sc ii}]$\lambda$5755 is detected in spectra of high-metallicity 
calibration H\,{\sc ii} regions, resulting in the determination of $t_2$. 
When [N\,{\sc ii}]$\lambda$5755 is detected in the spectrum of 
an H\,{\sc ii} region, then the intensity of [O\,{\sc iii}]$\lambda$4363 is estimated 
in the following manner. We calculate $t_3$ from the measured $t_2$ using  
the $t_2$ -- $t_3$ relation. We then estimate the line flux of  [O\,{\sc iii}]$\lambda$4363  
corresponding to the obtained value of t$_3$.

%====================================       Fig  No 6   OH - NO  mix
\begin{figure*}
\resizebox{1.00\hsize}{!}{\includegraphics[angle=000]{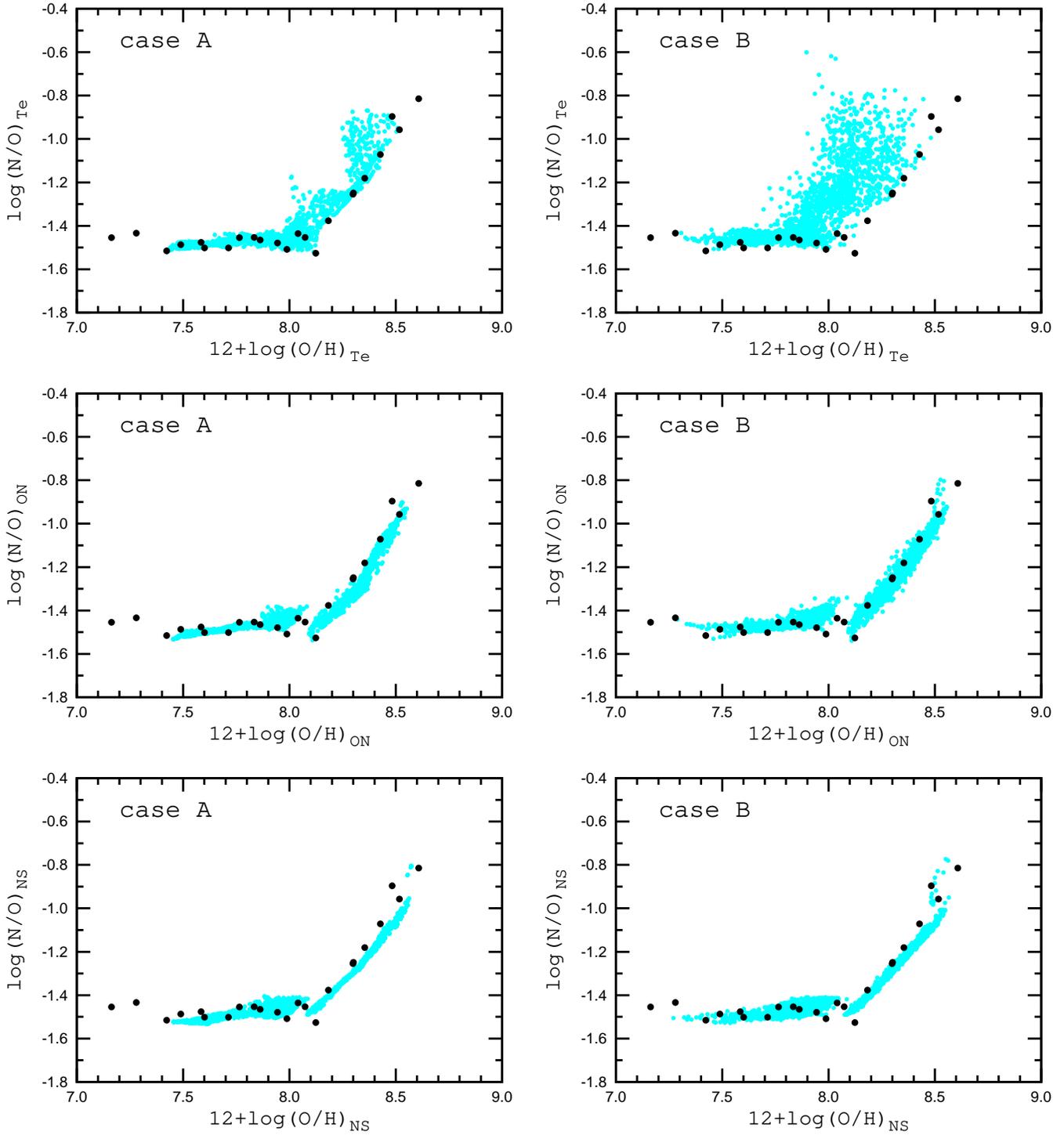}}
\caption{
The O/H -- N/O diagrams for the artificial composite nebulae
for case $A$ (left panels) and for case 
$B$  (right panels). 
The grey (light-blue) filled circles show the abundances derived in 
different ways from artificial global spectra. 
The dark (black) filled circles are $T_e$-based abundances in the 
components. 
(A color version of this figure is available in the online version.)
}
\label{figure:mix-ohno}
\end{figure*}

Using the calibration H\,{\sc ii} regions as components, we have performed a 
variety of Monte Carlo simulations, producing in each run an artificial 
spectrum of a composite nebula. 
The  calibration H\,{\sc ii} regions are not distributed uniformly within 
the considered metallicity range. We have thus divided the metallicity range in 
bins of $\Delta$(log(O/H)) = 0.05 and selected one calibration H\,{\sc ii} 
region within each bin according to the following two criteria:
\begin{enumerate}
\item The average value of the difference between the calibration-based and $T_e$ - based 
abundances is minimum. Both oxygen and nitrogen abundances and both ON and NS calibrations 
were taken into account in determining the average value of the difference.\\
\item The deviation of N/O abundances ratio from the N/O -- O/H relation 
 is not in excess of 0.15 dex. 
 \end{enumerate}
Fig.~\ref{figure:mix-mean} shows the positions of the 
selected calibration H\,{\sc ii} regions  
in the O/H -- N/O diagram by dark (black) filled circles.

At a fixed 12+log(O/H)$_{0}$,  we have simulated 100 global spectra, 
using the spectra of selected calibration H\,{\sc ii} regions with oxygen abundances 
within 12+log(O/H)$_{0}$ $\pm$ $\Delta$(log(O/H)). 
The contribution from each component to a given global spectrum of 
artificial composite  H\,{\sc ii} regions is defined by the value $w_{\rm j}$,  
where $w_{\rm j}$ are random numbers between -1 and 1. 
To take into account that some components may not contribute to a given 
spectrum we choose the interval of random numbers between -1 and 1 
instead of the standard interval between 0 and 1 and when $w_{\rm j} < 0$ we set it to 0, 
i.e. the component does not make contribution to the global spectrum. 
The component (calibration H\,{\sc ii} region) line fluxes are given on a scale where 
$F($H${\beta})=1$, therefore we can consider that the value $w_{\rm j}$ defines H${\beta}$ 
flux (or luminosity) of $j$ component in the given variant of the global spectra. 
Then the total $F($H${\beta})$ flux of the composite nebula is given by the expression: 
\begin{equation}
F({\rm H}{\beta})  =  \sum\limits_{{\rm j}=1}^{{\rm j}={\rm n}} w_{\rm j} , 
\label{equation:hbran}   
\end{equation}
where $n$ is the number of components in the composite nebula.

The flux $F($X$_{\lambda _{\rm k}})$ of the composite nebula in the 
line X$_{\lambda _{\rm k}}$ is given by:
\begin{equation}
F({\rm X}_{\lambda _{\rm k}})  = \frac{\sum\limits_{{\rm j}=1}^{{\rm j}={\rm n}} 
w_{\rm j} F_{\rm j}({\rm X}_{\lambda _{\rm k}})}
                       {\sum\limits_{{\rm j}=1}^{{\rm j}={\rm n}} w_{\rm j}}  .
\label{equation:ran}   
\end{equation}
We have computed the intensity of the 
H${\beta}$, 
[O\,{\sc ii}]$\lambda$3727, 
[O\,{\sc iii}]$\lambda$4363, 
[O\,{\sc iii}]$\lambda$4959, 
[O\,{\sc iii}]$\lambda$5007, 
[N\,{\sc ii}]$\lambda$6548, 
[N\,{\sc ii}]$\lambda$6584, 
[S\,{\sc ii}]$\lambda$6717, and 
[S\,{\sc ii}]$\lambda$6731 lines 
for the each artificial spectrum of composite nebulae.

The mean oxygen abundance of a composite H\,{\sc ii} region, 
weighted by the H$\beta$ line luminosity, can be determined as 
\begin{equation}
{\rm (O/H)}_{mean}  = \frac{\sum\limits_{{\rm j}=1}^{{\rm j}={\rm n}} 
w_{\rm j} {\rm (O/H)_j}}
                       {\sum\limits_{{\rm j}=1}^{{\rm j}={\rm n}} w_{\rm j}} .
\label{equation:ohm}   
\end{equation}
Similarly, the mean nitrogen abundance in a composite 
H\,{\sc ii} region is given by 
\begin{equation}
{\rm (N/H)}_{mean}  = \frac{\sum\limits_{{\rm j}=1}^{{\rm j}={\rm n}} 
w_{\rm j} {\rm (N/H)_j}}
                       {\sum\limits_{{\rm j}=1}^{{\rm j}={\rm n}} w_{\rm j}} .
\label{equation:nhm}   
\end{equation}

For each computed spectrum, we have estimated $t_3$,  
(O/H)$_{mean}$, (N/H)$_{mean}$, 
(O/H)$_{T_e}$, (N/H)$_{T_e}$, 
(O/H)$_{NS}$, (N/H)$_{NS}$, 
(O/H)$_{ON}$ and (N/H)$_{ON}$  abundances. 
This procedure was repeated for a variety of 12+log(O/H)$_{0}$ values .

\subsection{Global abundances in artificial composite nebulae}
%----------------------------------

Here, we discuss two cases, represented by two sets of Monte Carlo simulations:
\begin{itemize}
\item[{\bf A.}] We simulated 100 global spectra 
for each 12+log(O/H)$_{0}$, using components with oxygen abundances 
within 12+log(O/H)$_{0}$ $\pm$ 0.15, and varying 12+log(O/H)$_{0}$ 
from 7.5 to 8.5 in steps of 0.05 dex. 
The grey (light-blue) filled circles in the upper panel of Fig.~\ref{figure:mix-mean} 
show the mean abundances of the artificial composite 
nebulae for case $A$. The abundances of each component has 
been weighted by the corresponding H$\beta$-luminosities.\\

\item[{\bf B.}] We simulated 100 global spectra
for each 12+log(O/H)$_{0}$, using components with oxygen abundances 
within 12+log(O/H)$_{0}$ $\pm$ 0.35.  
The grey (light-blue) filled circles in the lower panel of Fig.~\ref{figure:mix-mean} 
show the mean abundances in artificial composite nebulae for case $B$. 
\end{itemize}

The N/O -- O/H diagrams for case $A$ with abundances obtained in 
different ways from the global spectra are shown in the left panels of 
Fig.~\ref{figure:mix-ohno} by grey (light-blue) 
filled circles. The upper panel shows the $T_e$-based abundances,
the middle panel the ON-calibration-based abundances,
and the lower panel the NS-calibration-based abundances in the 
artificial composite nebulae. 
The dark (black) filled circles in each panel are the $T_e$-based 
abundances of each component. 
The upper left panel shows that a significant fraction of the  
artificial composite nebulae with $T_e$ - based abundances are
shifted from the O/H -- N/O relation towards 
higher N/O ratios or/and towards lower oxygen abundances.

\begin{figure*}
\resizebox{1.00\hsize}{!}{\includegraphics[angle=000]{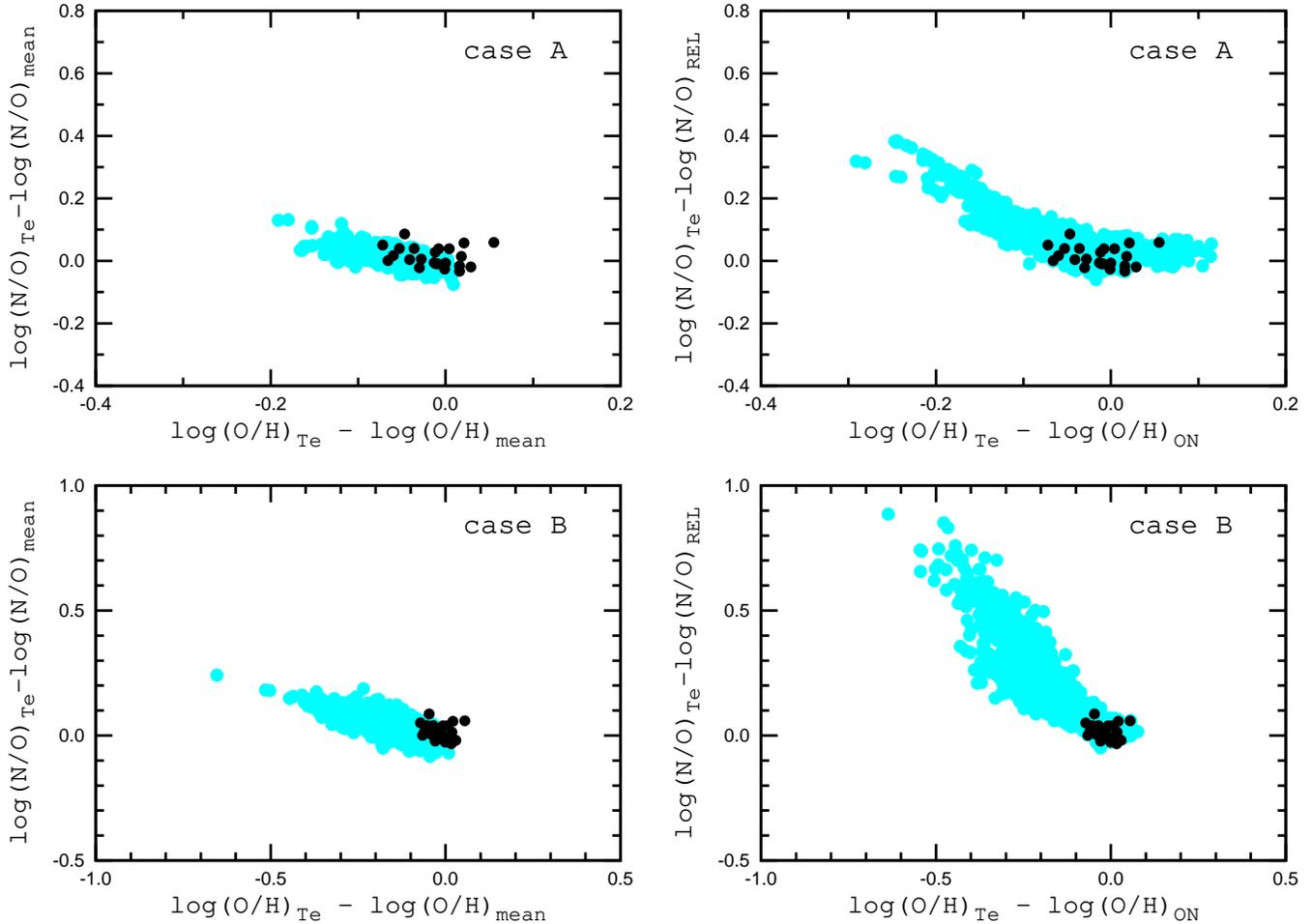}}
\caption{
{\it Left upper panel.} The grey (light-blue) filled circles show the difference 
between nitrogen-to-oxygen ratios (N/O)$_{T_e}$ and (N/O)$_{mean}$ in artificial composite nebulae 
as a function of the difference between oxygen abundances  (O/H)$_{T_e}$ -- (O/H)$_{mean}$
foe case $A$. 
The dark (black) filled circles show the components.
{\it Right upper panel.}  The grey (light-blue) filled circles show 
the deviation of (N/O)$_{T_e}$ ratios in artificial composite nebulae 
from the expected N/O -- O/H relation as a function of the difference between 
oxygen abundances derived with the $T_e$ method from global spectra and 
those obtained through ON calibration. 
{\it Lower panels} show the same but for case $B$. 
(A color version of this figure is available in the online version.)
}
\label{figure:mix-dno}
\end{figure*}

The left upper panel of Fig.~\ref{figure:mix-dno} shows the difference 
between nitrogen-to-oxygen ratios (N/O)$_{T_e}$ and (N/O)$_{mean}$ in the artificial 
composite nebulae as a function of the difference between oxygen abundances  
(O/H)$_{T_e}$ and (O/H)$_{mean}$ for case $A$ by the grey (light-blue) filled circles. 
The dark (black) filled circles show the components.
The right upper panel of Fig.~\ref{figure:mix-dno} shows the deviation of (N/O)$_{T_e}$  ratios 
in the artificial composite nebulae  from the N/O -- O/H relation 
(the linear fit discussed previously) as a function of the difference between the  
oxygen abundance derived by the $T_e$ method from global spectra and 
that obtained as the H$\beta$-luminosity-weighted mean abundances 
of components for case $A$.

%====================================       Fig  No 7   deviation dNO of mix

The N/O -- O/H diagram of $T_e$-based abundances for 
artificial composite nebulae for case $B$
is shown in the upper right panel of Fig.~\ref{figure:mix-ohno} by the dark (black) 
filled circles. The grey (light-blue) filled circles are  $T_e$-based abundances of the components. 
Again, a fraction of artificial composite nebulae with $T_e$-based abundances are  
shifted from the O/H -- N/O relation towards higher N/O 
ratios or/and towards lower oxygen abundances. 
The grey (light-blue) filled circles in the lower left panel of Fig.~\ref{figure:mix-dno} shows 
the difference between nitrogen-to-oxygen ratios (N/O)$_{T_e}$ and (N/O)$_{mean}$ in artificial 
composite nebulae as a function of the difference between oxygen abundances  
(O/H)$_{T_e}$ and (O/H)$_{mean}$ for case $B$. 
The dark (black) filled circles show the individual components.
The lower right panel of Fig.~\ref{figure:mix-dno} shows the deviations of (N/O)$_{T_e}$  ratios 
in the artificial composite nebulae from the N/O -- O/H relation 
(the linear fit discussed previously) as a function of the difference between the  
oxygen abundance derived by the $T_e$ method from global spectra and 
that obtained as the H$\beta$-luminosity-weighted mean abundances 
of components for case $B$.

By comparison of Fig.~\ref{figure:do-dno-s} and Fig.~\ref{figure:mix-dno} 
it is evident that artificial composite nebulae reproduce the anti-correlation between 
(log(N/O)$_{T_e}$ -- log(N/O)$_{REL}$) and (log(O/H)$_{T_e}$ -- log(O/H)$_{ON}$) 
observed in our sample of SDSS galaxies. 
This can be considered in favour of our hypothesis that the enhanced (N/O)$_{T_e}$ ratio 
derived for some SDSS objects can be due to the fact that 
these objects are composite nebulae, with  
a number of  H\,{\sc ii} regions with different 
physical properties contributing to the global spectrum.  
Hence, the electron temperature determined from the auroral-to-nebular lines ratio 
is overestimated, and the oxygen abundance is underestimated in such a nebula, 
i.e., it appears to be  a Peimbert temperature fluctuation effect (see the introduction).

The middle and lower panels of Fig.~\ref{figure:mix-ohno} show that 
both the ON and NS calibration-based abundances in the artificial composite 
nebulae follow, in general, the abundances of the components. 
However, both case A and B show a gap
near 12+log(O/H) $\sim$ 8.1. This gap is due to a transition between
two regimes in the calibrations, i.e.,  by the fact that both 
``warm'' and ``hot''  H\,{\sc ii} regions  \citep[according to the classification 
in][]{pilyugin2010ons}) make contributions to the global spectrum 
of composite nebulae. Therefore, such composite H\,{\sc ii} regions are neither purely 
warm nor purely hot. But distinct calibration relations for warm and hot
H\,{\sc ii} regions were in fact used \citep{pilyugin2010ons,pilyuginmattsson2011}. 
Hence, there appears to be a problem regarding how these calibration relations should be applied 
to such nebulae. It should also be noted that the criterion for distinguishing between 
warm and hot  H\,{\sc ii} regions is somewhat arbitrary.

We have performed a variety of Monte Carlo simulations, using different 
sets of components and different $\Delta$(log(O/H)) intervals, to  
produce artificial global spectra. We have found that the 
underestimation of the oxygen abundance 
with the $T_e$ method -- and, consequently, 
the shift of the position in the N/O -- O/H diagram of the 
artificial composite nebula with $T_e$-based abundances 
relative to its position with mean oxygen and nitrogen abundances --  
is larger for nebulae where the components have large differences in physical properties. 
In particular, the value of the shift increases with increasing $\Delta$(log(O/H). 
Thus, we have reached the following conclusions:
\begin{itemize}
\item If the composite nebula consists of H\,{\sc ii} regions with similar physical 
properties or a single  H\,{\sc ii} region which makes a dominant contribution 
to the global spectrum, then the oxygen and nitrogen abundances derived with the 
$T_e$ method and the ON and NS calibrations are in satisfactory agreement with  
each other, and near the mean H$\beta$ luminosity weighted oxygen and nitrogen  
abundances of the components in the composite nebula. \\ 
\item If H\,{\sc ii} regions with different physical properties make comparable 
contributions to the global spectrum of the composite nebula, then 
the $T_e$ - based oxygen abundance can be underestimated. Hence, the position of 
such a nebula in the N/O -- O/H diagram will be shifted towards lower oxygen 
abundances, mimicking an enhancement of the N/O ratio. \\
\item The ON and NS calibrations give oxygen and nitrogen abundances in the composite nebulae 
which agree with the mean luminosity-weighted abundances of their components 
to within  $\sim$0.2 dex. 
\end{itemize}

It should be noted that the high-metallicity  (12+log(O/H) $\ga$ 8.5) calibration H\,{\sc ii} regions 
with measured electron temperatures are very few. Thus, they cannot be 
simulated and we cannot investigate how similar (or dissimilar) their abundances estimated 
using the $T_e$ method and the strong-line calibrations are. 
Furthermore, our approach does not allow to take into account the 
contribution of a diffuse emission.

\section{DISCUSSION}
%======================================================

Our conclusion that the global abundances derived for composite nebulae with 
the classic $T_e$ method may be subject to systematic errors is not surprising. 
First, an effect of this kind has been predicted by  \citet{peimbert1967} for
H\,{\sc ii} regions with a non-uniform electron temperature. 
Second, \citet{kobulnickyetal1999} have compared physical conditions derived 
from spectroscopy of individual H\,{\sc ii} regions with those obtained  
from global spectroscopy for several galaxies with 12 + log(O/H)$\sim 8.2$, and
where the auroral line  [O\,{\sc iii}]$\lambda$4363 is detected 
in both individual and global spectra. They found that 
the standard  $T_e$ method using global spectra give systematic errors in that 
the oxygen abundances derived from global spectra are systematically 
0.05 -- 0.2 dex below the median values computed from spectra 
obtained in smaller apertures. 

The global oxygen abundances estimated with the one-dimensional $R_{23}$ calibration 
have been discussed by \citet{kobulnickyetal1999,moustakas2006}, while those obtained with the  
two dimensional  $R_{23}$  calibration ($P$  method) have been discussed by \citet{pilyuginetal2004a}.
It has been found that the abundance inferred from the integrated emission-line 
spectrum of a galaxy using one- or two-dimensional $R_{23}$  calibrations 
is representative of the gas-phase oxygen abundance at a fixed galactocentric radius 
(equal to 0.4 times the isophotal radius), as determined by the abundance gradient derived from H\,{\sc ii} regions. 
Here, we have found that the global abundances derived using the ON  and NS  calibrations 
are close to the mean abundances. Thus, the calibrations produce robust global abundances 
although both the absolute values of abundances in individual  H\,{\sc ii} regions and 
of global abundances depend on the adopted calibration.

%====================================       Fig  No 8    N2/R2 - t3  
\begin{figure}
\resizebox{1.00\hsize}{!}{\includegraphics[angle=000]{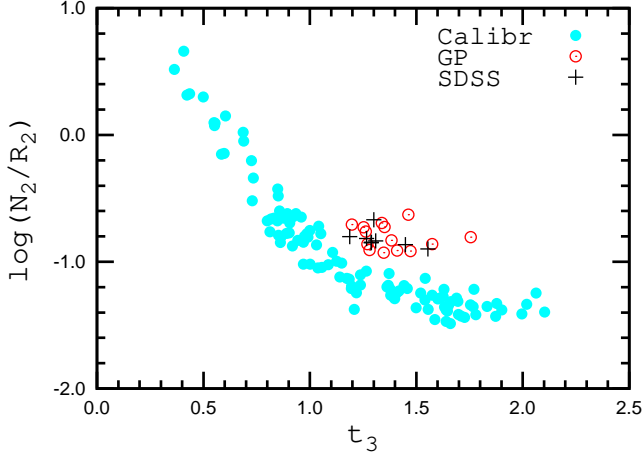}}
\caption{The flux ratio $N_2/R_2$ as a function of electron temperature $t_3$.
The grey (ligh-blue) filled circles show the calibration H\,{\sc ii} regions.
The SDSS objects with large ($>$ 0.3 dex) deviations from the expected N/O -- O/H relation are 
shown by dark (black) plus signs.  
The Green Pea galaxies with large deviations from the expected N/O -- O/H relation  
are shown by open (red) circles. 
(A color version of this figure is available in the online version.)
}
\label{figure:t-n2r2}
\end{figure}

%====================================       Fig  No  9    O/H - t3  
\begin{figure}
\resizebox{1.00\hsize}{!}{\includegraphics[angle=000]{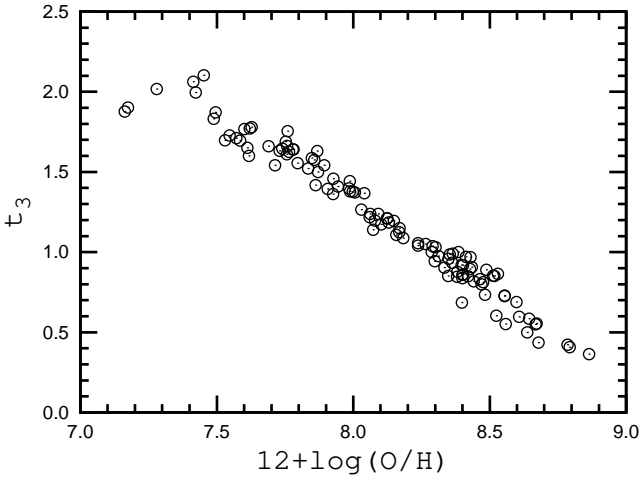}}
\caption{Electron temperature $t_3$ as a function of oxygen abundance
for the calibration H\,{\sc ii} regions.
}
\label{figure:zt-c}
\end{figure}

Our results suggest the $T_e$-based abundances in some composite 
nebulae can be incorrect and shift the positions of these nebulae 
in the O/H -- N/O diagram towards higher N/O abundances ratio or/and towards 
lower O/H abundances relatively to the positions of H\,{\sc ii} regions in 
nearby galaxies.  Thus the high (N/O)$_{T_e}$ abundances ratios obtained 
in some SDSS and Green Pea galaxies may be incorrect because they are composite 
nebulae. 

Fig.~\ref{figure:t-n2r2} shows the $N_2/R_2$ flux ratios as a function of the electron 
temperature $t_3$ for the calibration H\,{\sc ii} regions and for the SDSS and 
Green Pea galaxies with large ($>$ 0.3 dex) deviations from the expected N/O -- O/H relation. 
In the ON calibration, the $N_2/R_2$ flux ratios are used as temperature indicators. 
Fig.~\ref{figure:t-n2r2} shows that the measured $N_2/R_2$ in SDSS and Green 
Pea galaxies with $t_3$ $\sim$1.4 have values that are more typical for nebulae 
with $t_3$ $\sim$1. This means the two electron temperature indicators  
([O\,{\sc iii}]$\lambda$4959,$\lambda$5007/[O\,{\sc iii}]$\lambda$4363 
and $N_2/R_2$) are in conflict. Thus, to reproduce both the 
measured electron temperature $t_3$ and the $N_2/R_2$ flux ratios, the 
composite nebula should contain a component with a temperature $t_3$ $\sim$1 
as well as one with $t_3$ $\ga$ 1.4. 
The temperature difference between components can be larger than 5000 K, in some cases.
For example, the hottest Green Pea galaxy in our sample (with a  
large N/O deviation), J012910.15 + 145934.6, has $t_3$ = 1.75
while its measured $N_2/R_2$ flux ratio is typical for a nebula with $t_3$ $\sim$1.

Fig.~\ref{figure:zt-c} shows $t_3$ as a function of the oxygen abundance
12+log(O/H) for the calibration H\,{\sc ii} regions. 
We see that a H\,{\sc ii} region with $t_3$ $\sim$1 has 
12+log(O/H)  $\sim$8.3. 
Can a hot component of metallicity 12+log(O/H) $\sim$ 8.3 be responsible for  the 
high $t_3$ value ($\sim$ 1.4) in the composite nebula?

We can expect the points in the $t_3$ -- O/H diagram (Fig.~\ref{figure:zt-c})
to be close to the highest attainable electron temperatures $t_3$ for a 
given 12+log(O/H) because of a selection effect: electron temperatures 
are easiest to measure in the hottest H\,{\sc ii} regions. In such case, an    
H\,{\sc ii} region of metallicity log(O/H) = 8.3 cannot have an electron temperature 
as high as $t_3$ $\sim$ 1.4. Thus, a hot component with 12+log(O/H) $\sim$8.3 cannot 
be responsible for the high electron temperature measured.

According to Fig.~\ref{figure:zt-c}, a H\,{\sc ii} region with $t_3$ $\sim$1.4 has 
12+log(O/H)  $\sim$8.0. Furthermore, the $N_2/R_2$ flux ratio increases with decreasing  
electron temperature. Can a cold component with 12+log(O/H) $\sim$8.0 
be responsible for the high $N_2/R_2$? 

The typical N/O  ratio in  H\,{\sc ii} regions  with log(O/H) $\sim$ 8.0 is log(N/O) $\sim -1.5$.
A H\,{\sc ii} region with log(O/H) $\sim$ 8.0 and log(N/O) = --1.5 would have  
log$N_2/R_2 = -0.9$ at $t_3$ $\sim$0.5. However, the line 
fluxes $N_2$ and $R_2$ (normalised to the H$\beta$ flux) 
are one order of magnitude lower than the measured fluxes. 
Thus, a cold component of metallicity  12+log(O/H) $\sim$8.0 
cannot be responsible for the high $N_2/R_2$  flux ratio 
measured.

The measured high N/O ratios and high electron temperatures in some SDSS 
and Green Pea galaxies can be reproduced by composite nebulae involving 
components of different metallicities. However, the metallicity difference can  
be as large as a factor 2-3, or even larger in extreme cases. 
It is believed that Green Peas are low-mass galaxies 
\citep{cardamone2009,amorin2010,izotovetal2011}. 
The metallicity scatter of H\,{\sc ii} regions in typical dwarf galaxies is known to be 
small \citep[e.g.][]{croxall2009}.
This suggests that the measured high N/O ratios not in all the Green Pea galaxies 
can be reproduced by composite nebulae and some Green Pea galaxies may not be typical 
dwarfs galaxies.
A large metallicity scatter of the H\,{\sc ii} regions in a dwarf galaxy can arise 
if this galaxy has undergone an atypical evolution. 
For example, it can be the result of the merging of 
two dwarfs of different metallicities, triggering a starburst in both components 
simultaneously. Infall of non-enriched gas 
onto a well-evolved galaxy with a high N/O gas-phase ratio would 
decrease the oxygen and nitrogen abundances, and would result in a galaxy with 
an enhanced  N/O ratio. 
The evolution of a galaxy with selective heavy 
elements loss through enriched galactic winds can also result in 
enhanced  N/O ratios \citep[e.g.][]{pilyugin1993,yinetal2011}.

\section{CONCLUSIONS}
%====================

We have examined the oxygen and nitrogen abundances derived from global 
emission-line spectra of galaxies with 12 + log(O/H) ranging 
from $\sim$7.5 to $\sim$8.5, based on a sample of 281 SDSS galaxies 
with measured electron temperatures. 
The oxygen and nitrogen abundances in these
galaxies were derived with the 
$T_e$ method as well as with two recent strong-line calibrations: 
the ON and NS calibrations. 

For $T_e$-based abundances, the positions of some SDSS galaxies 
in the O/H -- N/O diagram are shifted towards higher N/O ratios 
or/and towards lower O/H abundances relative to the positions of H\,{\sc ii} 
regions in nearby galaxies. In case of strong-line calibration-based abundances, the   
SDSS galaxies occupy the same area  in the O/H -- N/O diagram 
as the H\,{\sc ii} regions in nearby galaxies. 

The global spectra of galaxies have been Monte Carlo simulated as a mix of 
spectra of individual components, based on well-observed H\,{\sc ii} regions 
in nearby galaxies. 
Abundance analysis of the artificial composite nebulae yields the following 
conclusions: 

\begin{enumerate}
\item If the composite nebula consists of H\,{\sc ii} regions with similar physical 
properties or a single  H\,{\sc ii} region makes a dominant contribution 
to the global spectrum, then the oxygen and nitrogen abundances derived with the 
$T_e$ method and the ON and NS calibrations are in satisfactory agreement 
with each other and near the mean H$\beta$ luminosity weighted  
value of oxygen and nitrogen abundances of the individual components of the composite nebula. 

\item If H\,{\sc ii} regions with different physical properties make comparable 
contributions to the global spectrum, then 
the $T_e$ - based oxygen abundances may be underestimated and the position of 
such a nebula in the N/O -- O/H diagram will be shifted towards lower oxygen 
abundances, mimicking an enhancement of the N/O ratio in the nebula. 
This effect is similar to the one discussed by Peimbert for  H\,{\sc ii} regions with 
small scale temperature fluctuations. For composite nebulae however,  this effect appears to be  
due to temperature variations on large spatial scales, caused by 
varying temperatures in different components of the composite nebula.  

\item The ON and NS calibrations give oxygen and nitrogen abundances in composite nebulae 
which agree with the mean luminosity-weighted abundances of their components to within  $\sim$0.2 dex. 
ON- and NS-calibration-based abundances for these nebulae also show a 
gap in abundance near 12+log(O/H) $\sim$ 8.1.

\item  The high (N/O)$_{T_e}$ ratios derived in some Green Pea galaxies
may be caused by the fact that their SDSS spectra are spectra of composite
nebulae made up of several components with different physical properties
(such as metallicity). However, for the hottest Green Pea galaxies, which
appear to be dwarf galaxies, this explanation does not seem to be
plausible.
It would work only if the HII regions in these galaxies have a dispersion
of abundances much larger than that typically found in dwarf galaxies.
\end{enumerate}

\section*{Acknowledgements}

We are grateful to the referee for his/her constructive comments.
L.S.P. acknowledges support from the Cosmomicrophysics project of
the National Academy of Sciences of Ukraine. 
J.M.V. and L.S.P. acknowledge the partial support of AYA2010-21887-C04-01 
from the Spanish PNAYA and CSD2006-00070 from CONSOLIDER 2010 programme 
of MICINN. 
L.S.P. thanks the hospitality of the Instituto de Astrof\'{\i}sica de 
Andaluc\'{\i}a where this investigation was carried out.
The Dark Cosmology Centre is funded by the Danish National Research Foundation.

Funding for the SDSS and SDSS-II has been provided by the Alfred P. Sloan Foundation, 
the Participating Institutions, the National Science Foundation, the U.S. Department of Energy,
the National Aeronautics and Space Administration, the Japanese Monbukagakusho, the Max Planck Society, 
and the Higher Education Funding Council for England. The SDSS Web Site is http://www.sdss.org/.

The SDSS is managed by the Astrophysical Research Consortium for the Participating Institutions. 
The Participating Institutions are the American Museum of Natural History, Astrophysical Institute Potsdam,
University of Basel, University of Cambridge, Case Western Reserve University, University of Chicago, 
Drexel University, Fermilab, the Institute for Advanced Study, the Japan Participation Group, 
Johns Hopkins University, the Joint Institute for Nuclear Astrophysics, the Kavli Institute for Particle 
Astrophysics and Cosmology, the Korean Scientist Group, the Chinese Academy of Sciences (LAMOST), 
Los Alamos National Laboratory, the Max-Planck-Institute for Astronomy (MPIA), the Max-Planck-Institute 
for Astrophysics (MPA), New Mexico State University, Ohio State University, University of Pittsburgh, 
University of Portsmouth, Princeton University, the United States Naval Observatory, and 
the University of Washington.


\begin{thebibliography}{}


\bibitem [Alloin et al. (1979)]{alloinetal1979} 
          Alloin D., Collin-Souffrin S., Joly M., Vigroux L., 1979, A\&A, 78, 200

\bibitem [Amor\'{i}n, P\'{e}rez-Montero \&  V\'{i}lchez (2010)]{amorin2010}
          Amor\'{i}n R.O., P\'{e}rez-Montero E.,  V\'{i}lchez J.M., 2010, ApJ, 715, L128 

\bibitem [Asari et al. (2007)]{asarietal2007} 
          Asari N.V., Cid Fernandes R., Stasi\'{n}ska G., Torres-Papaqui J.P., Mateus A., 
          Sodr\'{e} L., Schoenell W.,  
          2007, MNRAS, 381, 263

\bibitem [Bresolin, Kennicutt, \& Garnett (1999)]{bresolinetal1999}
          Bresolin F., Kennicutt R.C., Jr., Garnett D.R.,  
          1999, ApJ, 510, 104

\bibitem [Bresolin et al. (2005)]{bresolinetal2005}
          Bresolin F., Shaerer D., Gonz\'{a}lez Delgado R.M., Stasi\'{n}ska G.,  
          2005, A\&A, 441, 981

\bibitem [Bresolin (2007)]{bresolin2007}
          Bresolin F., 2007, ApJ, 656, 186

\bibitem [Bresolin et al. (2009)]{bresolinetal2009}
          Bresolin F., Gieren W., Kudritzki R.-P., Pietrzy\'{n}ski G., 
          Urbaneja M.A., Carraro G.,  2009, ApJ, 700, 309

\bibitem [Campbell, Terlevich \& Melnick (1986)]{campbell1986}  
          Campbell A., Terlevich R., Melnick J., 1986, MNRAS, 223, 811 

\bibitem [Cardamone et al. (2009)]{cardamone2009} 
          Cardamone C., Schawinski K., Sarzi M., et al., 2009, MNRAS, 399, 1191

\bibitem [Croxall et al. (2009)]{croxall2009} 
          Croxall K.V., van Zee L., Lee H., Skillman E.D., Lee J.C., 
          C\^{o}t\'{e}  S.,  Kennicutt  R.C., Jr. Miller B.W., 2009, ApJ, 705, 723 

\bibitem [Denicolo, Terlevich \& Terlevich (2002)]{denicoloetal2002}
          Denicolo G., Terlevich R., Terlevich E., 2002, MNRAS, 330, 69

\bibitem [Dopita \& Evans (1986)]{dopitaevans1986}
          Dopita M.A., Evans I.N., 1986, ApJ, 307, 431 

\bibitem [Erb et al. (2006)]{erbetal2006}
          Erb D.K., Shapley A.E., Pettini M., Steidel C.C., Reddy N.A., Adelberger K.L., 
          2006, ApJ, 644, 813

\bibitem [Ercolano et al. (2007)]{ercolanoetal2007}
          Ercolano B., Bastian N., Stasi\'{n}ska G., 2007, MNRAS, 379, 945

\bibitem [Ercolano et al. (2010)]{ercolanoetal2010}
          Ercolano B., Wesson R., \& Bastian N. 2010, MNRAS, 401, 1375

\bibitem [Garnett (1992)]{garnett1992} 
          Garnett D.R. 1992, AJ, 103, 1330

\bibitem [Izotov et al. (2006)]{izotovetal2006}
          Izotov Y.I., Stasi\'{n}ska G., Meynet G., Guseva N.G.,  
          Thuan T.X., 2006, A\&A, 448, 955

\bibitem [Izotov et al. (2011)]{izotovetal2011}
          Izotov Y.I., Guseva N.G., Thuan T.X., 2011, ApJ, 728, 161

\bibitem [Kennicutt (1992)]{kennicutt1992} 
          Kennicutt R.C., (Jr.), 1992, ApJ, 338, 310

\bibitem [Kniazev et al. (2004)]{kniazevetal2004}
          Kniazev A.Y., Pustilnik S.A., Grebel E.K., Lee H., 
          Pramskij A.G., 2004, ApJS, 153, 429 

\bibitem [Kobulnicky et al. (1999)]{kobulnickyetal1999} 
          Kobulnicky H.A., Kennicutt R.C., (Jr.), Pizagno J.L., 
          1999, ApJ, 514, 544

\bibitem [Liang et al.(2006)]{liangetal2006}  
          Liang Y.C., Yin S.Y., Hammer F., Deng L.C., Flores H., Zhang B., 
          2006, ApJ, 652, 257

\bibitem [McCall, Rybski \& Shields (1985)]{mccalletal1985}
          McCall M.L., Rybski P.M., Shields G.A., 1985, ApJS, 57, 1

\bibitem [Moustakas \& Kennicutt (2006)]{moustakas2006}
          Moustakas J., Kennicutt R.C., (Jr), 2006, ApJ, 651, 155 

\bibitem [Pagel et al. (1979)]{pageletal1979} 
          Pagel B.E.J., Edmunds M.G., Blackwell D.E., Chun M.S., Smith G., 1979, MNRAS, 189, 95

\bibitem [Peimbert (1967)]{peimbert1967}
          Peimbert M., 1967, ApJ, 150, 825 
 
\bibitem [Pettini \& Pagel (2004)]{pettinipagel2004}
          Pettini M., Pagel B.E.J., 2004, MNRAS, 348, 59L

\bibitem [P\'{e}rez-Montero \& Contini (2009)]{perezmontero2009}
          P\'{e}rez-Montero E., Contini T., 2009, MNRAS, 398, 949

\bibitem [Pilyugin (1993)]{pilyugin1993}
          Pilyugin L.S., 1993, A\&A, 277, 42  
		  
\bibitem [Pilyugin (2000)]{pilyugin2000}
          Pilyugin L.S., 2000, A\&A, 362, 325 
		  
\bibitem [Pilyugin (2001)]{pilyugin2001}
          Pilyugin L.S., 2001, A\&A, 369, 594

\bibitem [Pilyugin, Contini, \& V\'{\i}lchez (2004a)]{pilyuginetal2004a} 
          Pilyugin L.S., Contini T., V\'{\i}lchez J.M., 
          2004a, A\&A, 423, 427

\bibitem [Pilyugin, V\'{\i}lchez, \& Contini (2004b)]{pilyuginetal2004b} 
          Pilyugin L.S., V\'{\i}lchez J.M., Contini T.,  
          2004b, A\&A, 425, 849

\bibitem [Pilyugin \& Thuan (2005)]{pilyuginthuan2005} 
          Pilyugin L.S., Thuan T.X., 2005, ApJ, 631, 231

\bibitem [Pilyugin \& Thuan (2007)]{pilyuginthuan2007} 
          Pilyugin L.S., Thuan T.X., 2007, ApJ, 669, 290

\bibitem [Pilyugin et al. (2009)]{pilyuginetal2009} 
          Pilyugin L.S., Mattsson L., V\'{\i}lchez J.M., Cedr\'es B.,  
          2009, MNRAS, 398, 485

\bibitem [Pilyugin et al. (2010a)]{pilyuginetal2010a} 
          Pilyugin L.S.,  V\'{\i}lchez J.M.,  Cedr\'{e}s B., \& Thuan T.X.  
          2010a, MNRAS, 403, 896

\bibitem [Pilyugin et al. (2010b)]{pilyugin2010ons} 
          Pilyugin L.S., V\'{i}lchez J.M., Thuan T.X., 2010b, ApJ, 720, 1738

\bibitem [Pilyugin \& Mattsson (2011)]{pilyuginmattsson2011}
          Pilyugin L.S., Mattsson L.,  2011, MNRAS, 412, 1145

\bibitem [Stasi\'{n}ska (1978)]{stasinska1978}
          Stasi\'{n}ska G., 1978, A\&A, 66, 257 

\bibitem [Stasi\'{n}ska (2005)]{stasinska2005}
          Stasi\'{n}ska G., 2005, A\&A, 434, 507 

\bibitem [Stasi\'{n}ska (2006)]{stasinska2006}
          Stasi\'{n}ska G., 2006, A\&A, 454, L127

\bibitem [Stasi\'{n}ska (2010)]{stasinska2010}
          Stasi\'{n}ska G., 2010, Proceedings of the IAU Sump. No 262, 93

\bibitem [Thuan et al. (2010)]{thuanetal2010} 
          Thuan T.X., Pilyugin L.S., Zinchenko I.A.,  
          2010, ApJ , 712, 1029 

\bibitem [Tremonti et al. (2004)]{tremontietal2004} 
          Tremonti C.A., Heckman T.M., Kauffmann G., et al., 
          2004, ApJ, 613, 898

\bibitem [van Zee \& Haynes (2006)]{vanzeehaynes2006} 
          van Zee L., Haynes M.P., 2006, ApJ, 636, 214

\bibitem [van Zee et al. (1998)]{vanzeeetal1998} 
          van Zee L., Salzer J.J., Haynes M.P., O`Donoghiu A.A., 
          Balonek T.J.,  1998, AJ, 116, 2805

\bibitem [Vilchez \& Esteban (1996)]{vilchezesteban1996} 
          Vilchez J.M., Esteban C., 1996, MNRAS, 280, 720

\bibitem [Yin, Matteucci, \& Vladilo (2011)]{yinetal2011} 
          Yin J., Matteucci F., Vladilo G.,  
          2011, A\&A, 531, A136

\bibitem [York et al. (2000)]{yorketal2000} 
          York D.G., Anderson J.E., Anderson S.F., et al.,  
          2000, AJ, 120, 1579 

\bibitem [Zaritsky, Kennicutt \& Huchra (1994)]{zaritskyetal1994} 
          Zaritsky D., Kennicutt R.C., Huchra J.P., 
          1994, ApJ, 420, 87 


\end{thebibliography}
\end{document}